\documentclass[times, twoside]{zHenriquesLab-StyleBioRxiv-Preprint}
\usepackage{blindtext}
\usepackage{graphicx}
\usepackage{dblfloatfix} 
\usepackage{lineno}
\usepackage{comment}
\usepackage[normalem]{ulem}
\usepackage{color, colortbl}
\usepackage{MnSymbol}
\usepackage{tabularx}

\leadauthor{Häusler and Willomitzer} 

\begin{document}

\title{Reflections about holographic and non-holographic acquisition of surface topography}

\shorttitle{Reflections about holographic and non-holographic acquisition of surface topography}


\author[1,*]{Gerd Häusler}
\author[2]{Florian Willomitzer}

\affil[1]{University of Erlangen Nuremberg, Institute of Optics, Information, and Photonics,  91058 Erlangen, Germany}
\affil[2]{Northwestern University, Department of Electrical and Computer Engineering, 60208 Evanston, USA}
\affil[ ]{}

\affil[*]{Correspondence: \url{gerd.haeusler@fau.de}}

\maketitle

\begin{abstract}
\noindent
Recording and (computational) processing of complex wave fields offers a vast realm of new optical methods. Also for optical 3D-metrology. We discuss fundamental similarities and differences of holographic surface topography measurement versus non holographic principles, such as triangulation, classical interferometry, rough surface interferometry and slope measuring methods. Key features are the physical origin of the ultimate precision limit and how the topographic information is encoded and decoded. We demonstrate that the question “is holography just interferometry?” has different answers, depending on how we exploit holograms or interferograms for metrology. The answers will help users to find out if their measurement results could be improved or if they already hit the ultimate limit of what physics allows.
\end {abstract}

\vspace{-3mm}

\section{Introduction}
\noindent

\begin{table*}[t] 
\centering      
\begin{tabularx}{\textwidth}{| c | X | X | m{6.5cm} | X |}  
\hline\hline                        
\bf{Class} & \bf{Physical ~~~~~~~~~~~ Principle}  & \bf{Origin of meas. \ uncertainty $\delta z$} & \bf{Lower bound of meas. uncertainty $\delta z$} & \bf{Dependent on obs. aperture} \\ [0.5ex] 
\hline    \hline                
I & Triangulation & Speckle & $\delta z = C \lambda / (2 \pi \sin{u_{obs}} \sin{\theta})$ ~~~~~~~~~~~~~~~~or {$\delta z=~C \lambda / (2 \pi \sin^2(u_{obs}))$} ~~~~~~ for focus search & Yes \\    
\hline     
II & Classical ~~~~~~~~~~~~~~~ Interferometry & Photon Noise & principally, no lower physical bound & No \\ 
\hline     
III & Rough Surface Interferometry & Surface ~~~~~~~~~~~~~ Roughness  & Surface Roughness \ $  <|z-<z>|> \approx R_q$ & No \\ 
\hline     
IV  & Slope-Measuring Methods & Photon Noise & $\delta z \approx \delta x \cdot \delta \alpha \approx \lambda / {SNR}$ \ $\delta x = \lambda / \sin{u_{obs}} $&  Yes \\ [1ex]       
\hline     
\end{tabularx}
\caption{ Possibly all known sensors might fit into one of the four categories which differ in terms of the dominant source of noise and its dependence on the observation aperture. The table can be read as well from the right to the left, to find the correct class. $\lambda$ is the wavelength, $C$ is the speckle contrast ($C=1$ for laser illumination), $\sin{u_{obs}}$ is the observation aperture, $\theta$ is the triangulation angle (for focus searching methods, $\theta = u_{obs}$),  $\delta x$ is the lateral resolution, the roughness parameter $R_q$ is the standard deviation of the surface height $z$, $\delta \alpha$ is the slope uncertainty. $SNR$ is the signal-to-noise ratio.}
\label{tab:uncertainty}  
\end{table*}

For the first ever observers of a holographic recording, at the 1964 OSA spring meeting \cite{Leith_64}, the most intriguing feature was probably the three-dimensional appearance of the reconstructed object. After 60 years of holography this 'magical' effect still captivates observers who can freely change the viewing perspective and can locally focus onto the object surface. The reconstruction seems to be perfect, indistinguishable from the true object, including phase and amplitude. Holography is based on the interferometric superposition of waves, which suggests ultra precise measurement options, and indeed, holographic interferometry is a paradigm for this option \cite{Powell_65}, as well as the holographic Null test\cite{Offner_63} via computer generated holograms, which were invented by Adolf Lohmann \cite{Lohmann:67}. Holographic microscopy flourishes with the availability of high resolution camera chips \cite{Zhang:98, Kim_10, Kemper_07}. In this article we investigate holography and its major competitors in terms of what is similar and what is different. We focus on the acquisition of the (mainly macroscopic) surface topography in 3D space. 

A hologram is recording a complex field, sometimes called a "wave front" originating from an object under test. It can be read out optically or computationally. We can modify the recording by placing some optical instrument in front of the holographic plate, e.g, a shearing plate, for a proper source encoding. And we can place any optical instrument behind the (analog) hologram  to extract specific information about the object. With a camera chip replacing the hologram, any of those instruments and much more can be mimicked via computation.

In this article we are especially interested in the potential and the limits of measuring the surface topography $z(x,y)$, meaning the spatially resolved distance, via holographic methods. This topographical information can be deciphered by numerous methods discussed below. Our aim is to know the physical source of the dominant noise that ultimately limits the achievable precision - in other words the lowest possible statistical distance uncertainty $\delta z$. 

It will be discussed how holographic methods compare to the established non-holographic methods. Some methods have a close connection to holography, for others the connection is only indirect. We will give a look from the bird’s eye view. More details and further references can be found in \cite{Haeusler_11}. The idea of exploiting physical limits is described in \cite{Haeusler_19}. As a windfall profit we try to approach the FAQ “is holography just interferometry?” and give instructions for users to find out if his measurement results could be improved (for example) by better hardware - or if the results are already hitting the ultimate physical limit. The considerations below will also help the vendor of 3D-sensors to find out if the competitor can really satisfy the advertised specifications or if there might be some exaggeration.

To summarize: the features of our considerations are:
 \begin{itemize}
     \item{What is the physical origin of the ultimate measuring} uncertainty?
     \item{How is the topography encoded and decoded?}
     
 \end{itemize}

These questions are useful to bring some order into the overabundance of available 3D-sensors, and help customers to understand the limits and the potentials of different sensor principles. The questions avoid hardware aspects, just physical principles are considered.  

\begin{figure}[b!]
    \centering
    \includegraphics[width=1\linewidth]{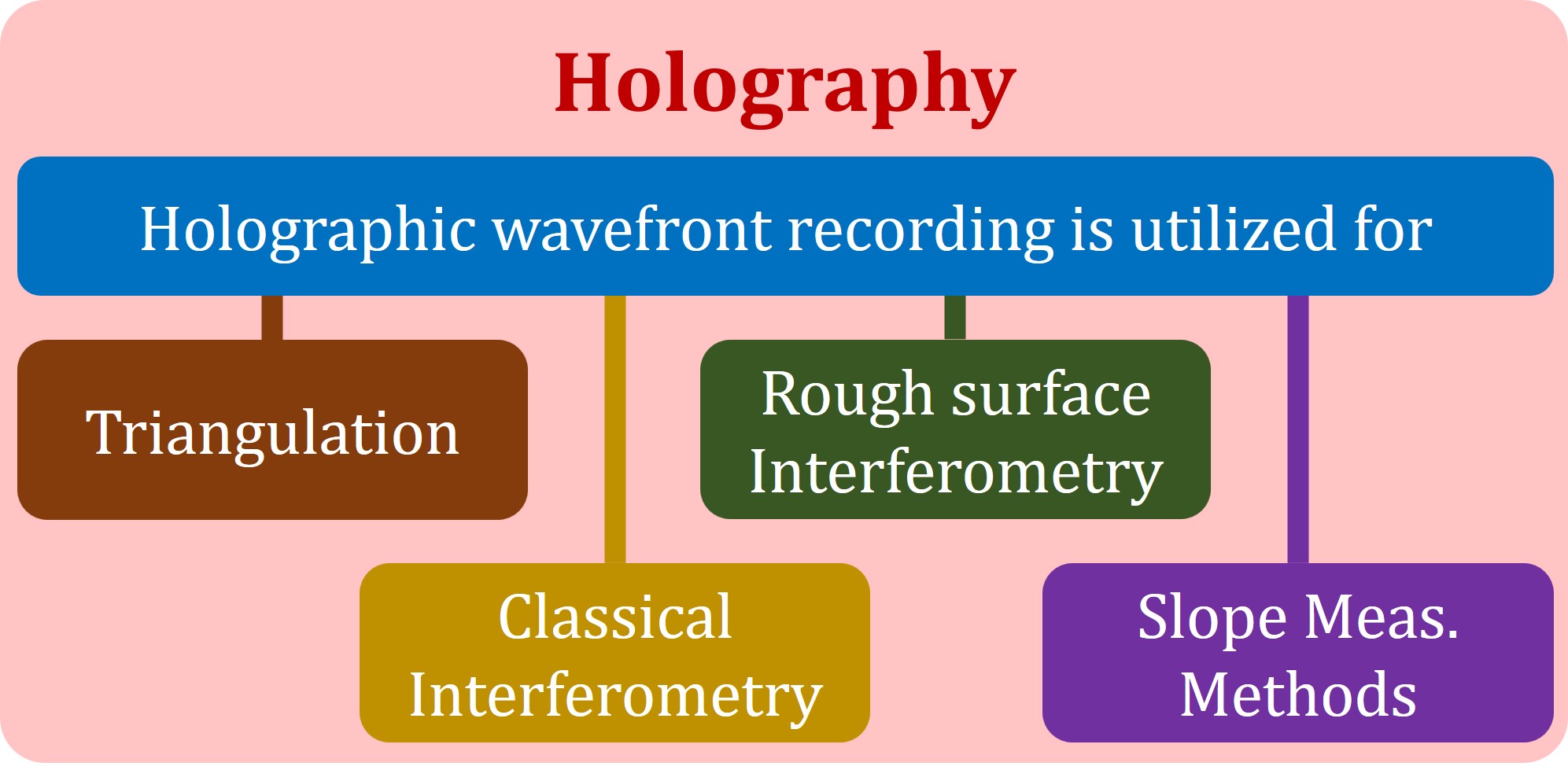}
    \caption{Holographic acquisition of surface topography: The oustanding feature to record an optical wavefront in a hologram can be utilized in many different ways to measure the surface topography of an object. Respective measurement principles  can be categorized into four groups: triangulation, classical interferometry, rough surface interferometry,  and slope-measuring methods.}
    \label{fig:HoloOverview}
\end{figure}

Optical sensors exploit different kinds of illumination, such as coherent/incoherent, structured/homogeneous, monochromatic/colored, polarized/unpolarized, … There are different ways of interaction with the object: coherent (Rayleigh scattering), incoherent (thermal, fluorescent), specular/diffuse, surface/volume… The detected modality may be the intensity, the complex amplitude, time-of-flight, polarization, coherence, …
Permuting all those parameters leads to more than 10,000 possible sensors. Not all of them are physically different, but there are serious differences which are significant for the user of 3D-sensors. It turns out that all sensors (as far as considered by the authors) belong to four different measuring principles which can be classified with respect to the physical origin of the statistical measuring error, or “precision”. A further parameter, important for users, is the dependence of the precision (here not the accuracy and lateral resolution) on the observation aperture. We refer to earlier investigations \cite{Haeusler_11}. The results are condensed in Tab.~\ref{tab:uncertainty}. The competing measuring principles are: triangulation (I), classical interferometry (II), rough surface interferometry (III). There is a category IV, we name it "slope-measuring methods". This class comprises methods that intrinsically measure the surface slope or lateral derivative. Table~\ref{tab:uncertainty} summarizes the fundamental distinction between the categories or classes and displays the recipes to calculate the limits. The Figs. ~\ref{fig:Mattscheibe}, \ref{fig:Nozzle}, \ref{fig:Defl}  illustrate the physical background leading to the ultimate source of measuring uncertainty.

A few examples for each class are given: Class I, triangulation, comprises amongst others: laser triangulation, stereo-photogrammetry, focus search, fringe projection triangulation, structured illumination microscopy (The wide area of fluorescence methods is not discussed here).  Class II, classical interferometry: this is essentially interferometry at specular surfaces, where the local surface topography $z(x,y)$ within the diffraction disc varies in a way that $ exp(ikz) \approx 1+ i kz(x,y)$, so we do not see speckles, and coherent lateral averaging over this area leads to an averaged distance $<z(x,y)>$, instead of $<exp(ikz(x,y)>$ which is non-linear and non-monotonic ('chaotic')  in $z$.  In class III we find rough surface interferometry. It includes rough-surface two-wavelength interferometry and rough-surface scanning white-light interferometry (“coherence radar”). In class IV we find incoherent methods such as phase measuring deflectometry, the Hartmann-Shack sensor, photometric stereo and as well coherent methods which exploit shearing interferometry with its modifications, such as differential interference contrast.

Until today the authors did not find optical 3D sensors that does not fit within one of these categories (the authors encourage the reader to find such sensors). Now we are ready for the question: where do we find holography? Our fast answer is: it depends on the application.

\section{Holography vs. triangulation}
\label{sec:Tria}

\noindent

\begin{figure}[t!]
    \centering
    \includegraphics[width=1\linewidth]{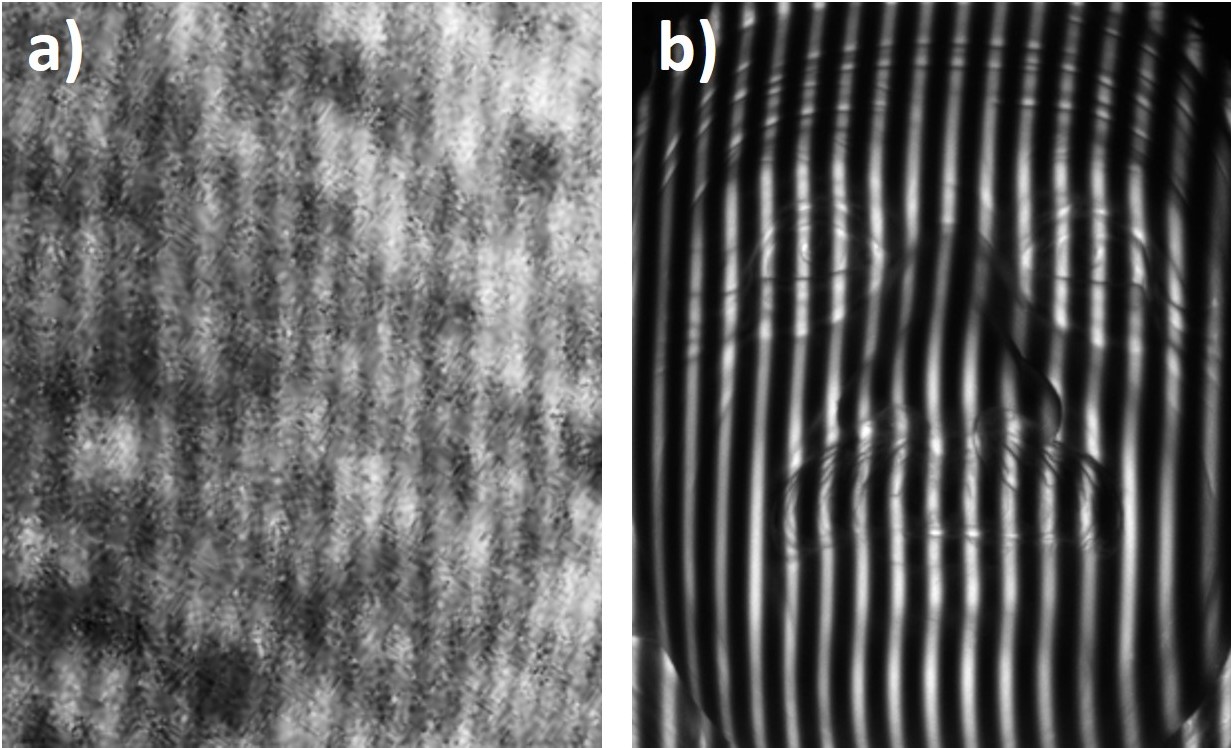}
    \caption{Systems-theoretical similarity between holographic encoding and encoding in fringe projection triangulation.  Both images can be decoded by single-sideband demodulation. a) Close-up photograph of a transmission hologram which was recorded on photographic emulsion \cite{HoloImage}. b) Fringe projection triangulation camera image for the measurement of the face of a plaster bust. }
    \label{fig:HoloFringe}
\end{figure}

Figure~\ref{fig:HoloFringe} displays a striking similarity between the appearance of a hologram of a rough object  and the camera images taken for line- or fringe projection triangulation \cite{Srinivasan_84, Takeda_83, Willo_17, Huber_13}: both display fringes with a carrier frequency $f_0$ where the local object depth $z(x,y)$ is encoded by the fringe phase.

Even more, the decoding of a hologram is information-theoretically the same as the decoding process in single-shot fringe projection triangulation (so-called “Fourier transform profilometry” \cite{Takeda_82, Takeda_83}). Both are decoded by single-sideband demodulation, respectively by optical separation of the zero order from the $+/- 1^{st}$ diffraction orders. And both methods share the same space-bandwidth limitations: only one third of the available space-bandwidth can be used \cite{GH_12, Willo_17} which is a big challenge specifically (but not only) for digital holography. For static objects, 2/3 of the expensive space bandwidth can be replaced with temporal bandwidth via phase shifting with at least three subsequent exposures - which is now standard for many sensors based on incoherent illumination, interferometry or holography \cite{Bruning_74}. For fringe projection triangulation, there is one more serious problem, the ambiguity- or indexing problem. For static scenes it can be solved by the acquisition of a second image sequence with another fringe frequency.

Single-shot measurements are possible, but not with a dense 3D point cloud, for a simple reason: One camera pixel cannot deliver the necessary information about the local illumination, the local reflectivity and the distance $z(x,y)$ of an object point.

Typically, single-shot sensors display point cloud densities much lower than 1/3 \cite{Willo_19Diss}  as the available space bandwidth additionally needs to account for the solution of the indexing problem. The “single-shot 3D movie camera” \cite{Willo_19Diss, Willo_17, Willo_15} solves this problem with a trick that (from a systems-theoretical side) shows strong similarities to two-wavelength holography or -interferometry: Instead of a sequence of images taken at different wavelengths (or fringe-frequencies), two cameras simultaneously capture two images of the fringe-encoded surface from two different viewing angles. This is noteworthy, as the mathematical structure of the deocding algorithms strongly resemble those used in dual-wavelength interferometry \cite{Cheng_84, Falaggis_11, Willo_19Diss}.

So far about the systems theory of encoding and decoding. Irrespective of the system theoretical similarities, the physics of the encoding is fundamentally different in both principles. In a hologram it is the optical path length together with the reference wave that encodes the (interference) fringe phase, respectively the local distance $z(x,y)$. In the triangulation system it is just the perspective distortion of the fringe pattern that is projected onto the object surface, see Fig.~\ref{fig:HoloFringe}

\begin{figure*}[b!]
    \centering
    \includegraphics[width=1\linewidth]{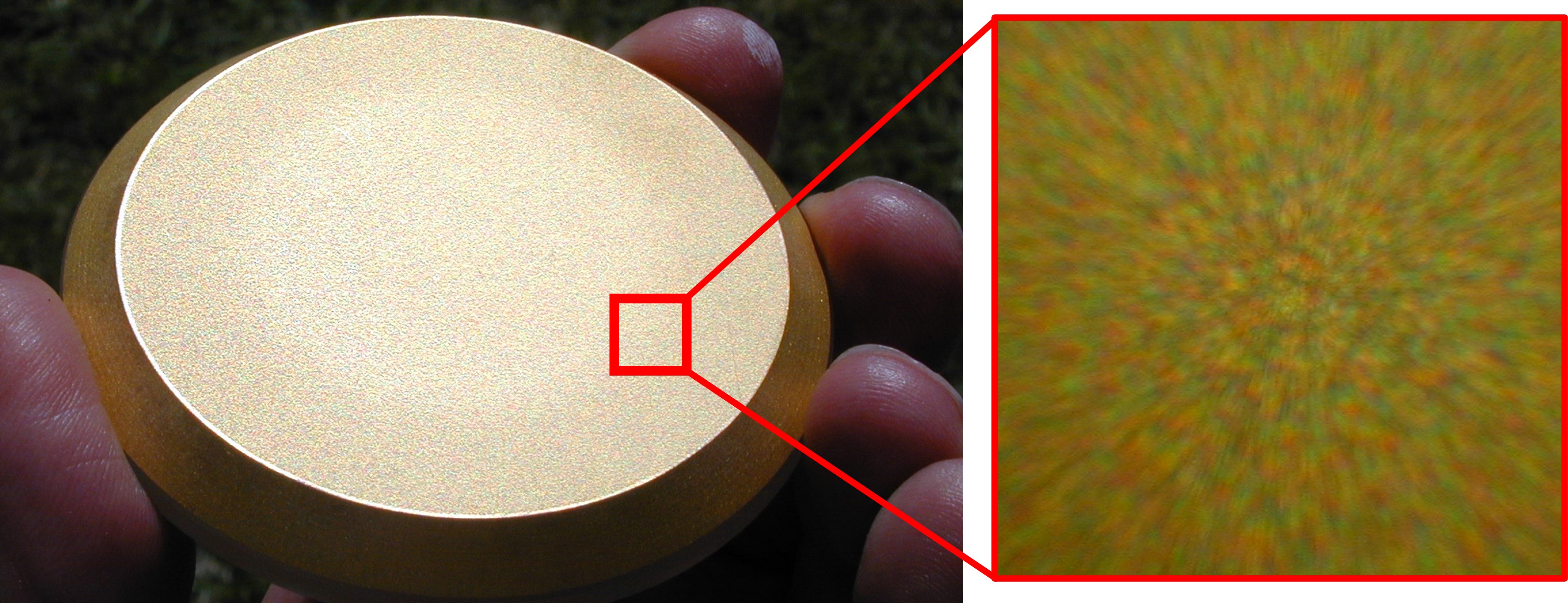}
    \caption{Groundglass in sunlight: at close distance - with the observation aperture angle close to the illumination aperture angle - speckles can be seen even with a white extended light source}
    \label{fig:Mattscheibe}
\end{figure*}

\begin{figure}[b!]
    \centering
    \includegraphics[width=1\linewidth]{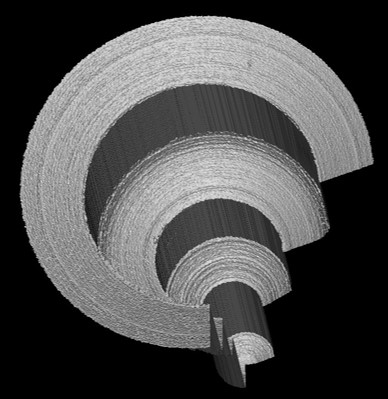}
    \caption{Gauge for an injection nozzle (about 10 mm long), measured by coherence radar. In spite of the low aperture that allows for measuring within deep holes, the measuring uncertainty is a few micrometers only.}
    \label{fig:Nozzle}
\end{figure}

\begin{figure*}[t!]
    \centering
    \includegraphics[width=1\linewidth]{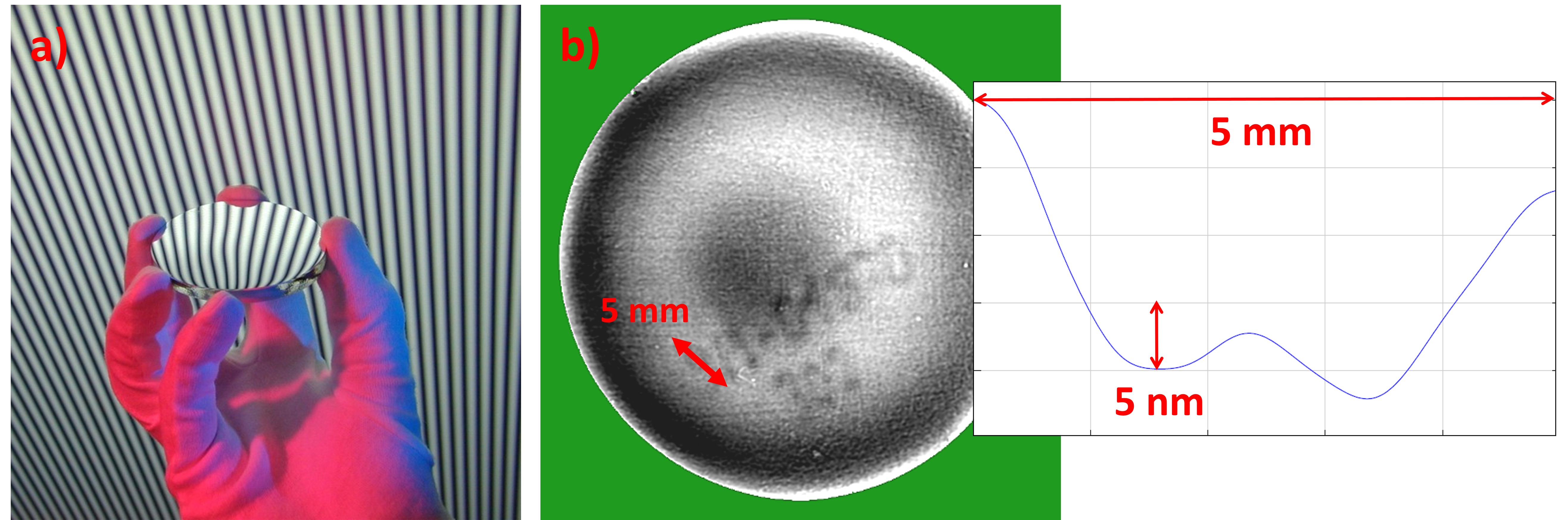}
    \caption{Phase measuring deflectometry. (a) basic principle, (b) "ghost writing": A lens was marked with a white board marker, later the numbers were erased. The few nanometer damage of the lens surface can be detected and quantitatively measured. }
    \label{fig:Defl}
\end{figure*}

For rough objects, a perfect wave front reconstruction was impossible before Gabor, Leith and Upatnieks. Classical interferometry does not work because of the random speckle phase at the hologram plane as well as at the image plane.   It was just the off-axis reference wave that eventually encoded the speckle phase in a way to allow the object wave being easily separated from other disturbing terms (volume holograms exploit additional 3D space-bandwidth for storage which opens new options not discussed here). The origin of speckles is surface roughness. The holographic reconstruction of the object (optically or by numeric back projection) will also display speckles. Watching the seemingly noise-free triangulation image of Fig. 2b\ref{fig:HoloFringe} more carefully, it turns out that the fringe phase displays a small random error as well. This error is caused by the partial spatial coherence of the illumination in combination with observation. There is always some residual spatial coherence, even with a large white incoherent light source (!), which results in an ultimate lower limit of the measuring uncertainty $\delta z$ for the local distance $z$ of an object point \cite{Dorsch_94}. 
Figure~\ref{fig:Mattscheibe}, illustrates the 'ubiquituous spatial coherence'.  Already partial spatial coherence disturbs measurements of rough surfaces, even if speckles are being unnoticed by a distant observer (with a small observation aperture).

\begin{equation}
    \delta z = \frac{C}{2\pi} \frac{\lambda}{\sin{u_{obs}} \sin{\theta}} ,
    \label{eq:dz_theta}
\end{equation}

where $\theta$ is the triangulation angle, $ \sin{u_{obs}}$ is the observation aperture and $C$ is the speckle contrast ($C=1$ for laser illumination, so for most holograms). For focus-searching principles Eq.~\ref{eq:dz_theta} degenerates to the Rayleigh depth of field (neglecting the C/2$\pi$ factor),  as $u_{obs} $ acts as effective triangulation angle:

\begin{equation}
    \delta z = \frac{C}{2\pi} \frac{\lambda}{\sin^2{u_{obs}}} 
    \label{eq:dz_u}
\end{equation}

For spatially incoherent light sources, the contribution $C_s$ of spatial coherence to the speckle contrast $C$ can easily be estimated by the observation aperture and the illumination aperture \cite{GH_04_Enc} (for the other contributing factors polarization, temporal coherence and averaging via large pixels see \cite{GH_04_Enc, Dorsch_94, Willo_19Diss}):

\begin{equation}
   C_s = \min{(\frac{\sin{u_{obs}}}{\sin{u_{ill}}} , 1)}
    \label{eq:Cs}
\end{equation}

An illustrative daily life example: for laser triangulation (with $C=1$, $\sin{u_{obs}} = 0.01$, $\sin{\theta} = 0.2$) we achieve a precision of about $40 \mu m$. With spatially ‘incoherent’ fringe projection using a large incoherent illumination aperture (see Eq.~\ref{eq:Cs}), the same specifications may lead to a precision about four times better. 

These beneficial properties made fringe projection triangulation a well established ‘gold standard method’ for many macroscopic applications.  A well designed sensor avoiding as much spatial coherence as possible may display a dynamical range of up to 10,000 depth steps. We emphasize that efficient metrology is enabled only by efficient source encoding. In optical metrology it is the proper \textit{\textbf{illumination}} that encodes the depth in an information efficient way (with low redundancy) \cite{Wagner_03}.

Coming back to holography: Can we decipher the surface topography from a hologram? At a first glance, this should not be a problem, as we can “see” the 3D reconstruction (this marvelling feature can hopefully be exploited for displays, in the future). But what about exploitation for metrology? One could suspect that the object surface can be fully reconstructed, within the limits of diffraction theory, where the size of the hologram can be seen as the limiting aperture. We approach the answer via an extreme counter-example: a hologram of a diffusely scattering white planar surface. It is impossible to find the surface by focusing through the hologram, as it is impossible to focus a camera onto a white wall in daylight. The deep reason is that from the hologram the wave field in front of the holographic plate can be restored, but (generally) we cannot localize the millions of surface points where the individual spherical waves have been scattered in the direction to the holographic plate. It will be discussed later that holography can access not just a wave front but after all even the coherence function and the surface topography.

To have access to the surface from a simple hologram, some structure is necessary, either via inherent features (salient points), or via structured illumination as used in fringe projection triangulation. This is valid for the holographic reconstruction and as well for incoherent methods. Summarizing, we have to admit that although holography - as interferometry - encodes distance via the phase of the propagating waves - deciphering of the surface topography is not possible by the first, without further means. It should be added however that classical interferometry commonly looks at the image plane, so there is a priori information about the location of the object. Principally, interferometric measurements can measure a wave front, but not the true surface of a remote object, without further information.

We illustrate this by an example: from the image of a laser spot that is projected onto a rough object, the distance z can be found via focus search or shearing interferometry, in spite of speckle \cite{GH_88, GH_88_S}. However, it was shown that both methods have to be attributed to class I, which possibly might surprise the reader. The explanation follows in section  \ref{sec:slope} about slope measuring methods.

Triangulation of a rough surface with laser illumination suffers from a serious measuring uncertainty, determined essentially by the observation aperture: These properties can be looked up in Tab.~\ref{tab:uncertainty}. It is obvious that triangulation measurements through a hologram, and via indirect measurements by numeric back propagation must be attributed to class I as well. There is one exception:  if the object displays fluorescence or is thermally excited, then there is no speckle (and no hologram) because of perfect spatial incoherence \cite{Thorley_14, spellenberg_01}. Microscopy based on fluorescence microscopy can localize molecules better than nanometers. 

So the ultimate precision of holographic (rough surface) topography measurement via focus search (or related methods) is given by Eqs.~\ref{eq:dz_theta} and ~\ref{eq:dz_u}, and not by the much lower photon noise of classical interferometry. As a consequence, the precision of triangulation strongly depends on the observation aperture.  Fortunately, by exploiting the large aperture that holography may easily provide, the precision can be comparably high, in spite of speckle noise.

\section{Holography vs. rough surface interferometry}
\label{sec:RoughInt}

As mentioned, the problem of phase randomization in a speckle pattern forbids the topography measurement of rough surfaces with classical single-wavelength interferometry. A solution which is well established now is  rough-surface scanning white-light interferometry ('coherence radar') \cite{Dresel92, Caber93}.

A partial solution that can be implemented by holography, is two-wavelength holography, enabling the acquisition of the topography by contouring: A hologram of the surface is made by illuminating the object and the holographic plate with a wavelength $\lambda_1$. In a second step the holographic plate as well as the object are illuminated with a wavelength $\lambda_2$, slightly different from $\lambda_1$. The object recorded at $\lambda_1$ is played back with $\lambda_2$ and superimposed with the  waves from the real object illuminated at $\lambda_2$. The interference pattern displays contour lines at a distance given by the “synthetic” wavelength \cite{fercher85, vry86, dandliker88, Li18}

\begin{equation}
   \Lambda = \frac{\lambda_1 \lambda_2}{|\lambda_1 - \lambda_2|}
    \label{eq:SWL}
\end{equation}

which is the beat wavelength between $\lambda_1$ and $\lambda_2$ and hence can be picked orders of magnitudes larger. The light at the optical wavelengths $\lambda_1$ and $\lambda_2$ acts as a carrier for the synthetic wave. This is important because the object serves as a “\textit{rough} mirror”, in spite of the large synthetic wavelength. Incoming light is scattered and the object can be seen from all directions. At the same time, the observed interference fringes at the synthetic wavelength do not display phase randomization. It is noteworthy that even a surface without features like edges or texture can be reconstructed. Bright contour lines are found just at the location where the two waves (with $\lambda_1$, $\lambda_2$)  display the same phase.  For two-wavelength holography the temporal coherence function is periodic, causing an ambiguous reconstruction. This can be avoided by 'multi-wavelength' holography or white-light interferometry \cite{Wagner00, Dresel92, Willo21}. 

This remarkable advantage of two-wavelength interferometry and -holography has been utilized recently for the relatively new research field of “Non-Line-of-Sight” imaging, which is concerned with the task of looking around corners and imaging through scattering media. To look around a corner, a synthetic wavelength hologram of an object obscured from direct view is captured. The hologram is taken by imaging a remote surface (such as a wall), which can “see” both the hidden object and the sensor unit. This remote surface acts as a “virtual holographic plate” and the object is reconstructed by holographic back-propagation of the synthetic hologram at the synthetic wavelength \cite{Willo21, Willo19COSI, Willo19Arx}. 

After understanding the role of roughness qualitatively, again the obligatory question: How to categorize two-wavelength holography in terms of noise? The resulting images with more or less ‘un-speckled’ fringes suggest an attribution to “classical interferometry”, which is characterized by photon noise as the cause of the precision limit. Combining the advantages of diffuse scattering with visible light illumination,  without the drawbacks of speckle, sounds like magic, meaning that carefulness is advisable. And indeed, two-wavelength holography is NOT attributed to the same class as classical interferometry.  Two-wavelength holography and -interferometry belong to class III, where the precision is ultimately limited by the surface roughness: 
 
The reason is that the speckles produced by the two closely spaced wavelengths $\lambda_1$, $\lambda_2$ display a small phase decorrelation \cite{vry86, Willo21, Li21} resulting in a random phase difference $d\varphi_{1,2}$. This phase difference may correspond to a path difference of only $ ~ \lambda/100$ (for example). But in the output signal, the measured path difference is „magnified“ by $\Lambda / \lambda$ resulting in a random distance error of $\Lambda/100$, instead of $ \lambda/100$, in this example. This measuring uncertainty is much larger than the shot noise limit of classical interferometry. 

We now estimate the random phase error and its physical cause. We assume that the two speckle patterns at $\lambda_1$ and $\lambda_2$ are sufficiently spatially correlated, as we select  the synthetic wavelength  to be much larger then the surface roughness \cite{George75, Willo21}.  At a certain image point $(x’, y’)$, waves are accumulated from a small object area given by the back projected diffraction point spread function of the observing lens. This area will be approximately the size of a back projected subjective speckle. Within this area, the rough surface will have a ‘summit’ or peak and there will be a deepest valley point. The peak-valley distance $R_t$ (over a certain length) is one of the common roughness parameters.  $R_t$ determines the maximum possible phase difference $d\varphi_{1,2}$ between the waves accumulated at $(x’, y’)$. For the two wavelengths $\lambda_1$, $\lambda_2$, we easily find:

\begin{equation}
  d\varphi_{1,2}  = 2 \cdot (\frac{2 \pi R_t}{\lambda_1} -\frac{2 \pi R_t}{\lambda_2} ) = \frac{4 \pi R_t}{\Lambda_1}
\end{equation}

And with $\delta z = \Lambda d\varphi_{1,2} / 2 \pi$  we get the measuring uncertainty at the position $(x’, y’)$:

\begin{equation}
    \delta z = 2 R_t
    \label{eq:deltaz_rough}
\end{equation}

Equation~\ref{eq:deltaz_rough} tells us that the phase decorrelation between the two wavelengths leads to a limit of the measuring uncertainty, given just by the surface roughness. Although an approximation, the physical cause of the measuring uncertainty is found. It is the same cause, even with a very similar quantitative result, that was found in a rigorous analysis for scanning white-light interferometry at rough surfaces \cite{Ettl95, GH99Asa}. 

A few more clarifying words about rough surface white-light interferometry ('coherence radar') \cite{Dresel92, Caber93}:  

The micro-topography of the most "rough" surfaces such as a white wall, a ground glass or a machined surface cannot laterally be resolved, unless we use a high aperture microscope. But even with a small observation aperture, within deep boreholes and from a large distance, we can measure the topography of rough surfaces, without being able to laterally resolve the micro-topography.  The statistical measurement error (the precision) $\delta z$ is always given just by the surface roughness. As most daily life surfaces or technical objects have a roughness of only a few micrometers, the user of such instruments will get a really small  measuring uncertainty much better than by triangulation.  And, as a further present, this is independent from the observation aperture.

With the benefit of hindsight, the similarity of the physical cause of the dominating noise is not surprising, as both methods, white light interferometry and two-wavelength holography exploit an interplay of two (or more) wavelengths. The latter measures the surface topography via the synthetic wavelength $\Lambda $, which displays a phase uncertainty given by the surface roughness, as explained above.

We summarize this section by assigning two-wavelength holography and two-wavelength interferometry to class III: As these methods exploit just 'time-of-flight' or optical path length, the ultimate limit of height precision is given by the surface roughness and does not depend on the observation aperture.  \cite{GH99Asa, Haeusler_11}.

Again: the ultimate precision of rough-surface \textbf{scanning white-light interferometry} is only determined by the object and not determined by the instrument, which is a marvelling near magic physical and information theoretical peculiarity. However, the ultimate limit of rough-surface\textbf{ two-wavelength interferometry} displays the same limit, in spite of the large synthetic wavelength. This is somewhat disappointing and unexpected, perhaps even for the expert.

A short appendix:  It is well known that illumination by a wavelength significantly larger than the surface roughness helps to measure rough surfaces. One could use far infrared light \cite{kwon80}, but this creates a new problem: besides technical issues and the limited availability of infrared detectors, the surface now acts as a mirror. For 3D objects strongly deviating from a plane, light will scarcely find its way back to the hologram plate/detector.  It is important to note that the physical properties of such an experiment must be looked up under category II, “classical interferometry” as the surface is smooth for the long wavelength. We see that changing the illumination may change the physical background dramatically.

\section{Holography vs. slope measuring methods} \label{sec:slope}

The so far discussed methods intrinsically measure either the lateral position of a local feature (which is translated into distance via triangulation), or they measure the distance via an interferometric measurement of the phase (or time-of-flight) of a coherent (or partially coherent) signal. However, there also exists a class of physically completely different methods - where the intrinsic signal is the local slope. ‘Intrinsic’ means that the slope is not evaluated by a-posteriori differentiation – the slope is already encoded in the optical signal before arriving at the photodetector. This is an invaluable feature, as the encoding is done optically before the detector noise is added. Hence, some of these methods enable sub-nanometer precision for local surface height variation, by very simple means. This is because (well designed) slope measurement methods exploit a perfect source encoding: the OTF represents the spatial derivative (or at least an approximation,for shearing methods). 
Among the spatially differentiating methods we find the so called phase-measuring deflectometry (pmd) \cite{GHDefPat, Knauer04, Willo20Def} which is completely incoherent, and we find classical shearing interferometry \cite{Bates46}.

Both methods measure specular surfaces or wave-fronts. For rough surfaces, shearing holography comes into play, specifically because of its unlimited options for post processing of the holographic data.  We do not discuss  the so called “photometric stereo” here, as it requires Lambert scattering, it is based on accurate intensity measurements and hence, and it is sensitive against unavoidable spatial coherence (small sources to generate shading).

As starters, a brief explanation of incoherent phase-measuring deflectometry for a later comparison with holography and other coherent methods \cite{GHDefPat, Knauer04, Willo20Def}:

The simple idea: A large screen with an incoherently radiating sinusoidal fringe pattern is in remote distance from the specular object under test. The observer watches the screen that is mirrored by the object. If the object is not planar,  the captured fringes in the camera image are distorted.

From the distortion, measured by phase shifting, local surface deformations even below the nanometer range can be measured. There is as well a microscopic realization \cite{GH08MD}, where weak phase objects can be measured in transmission \cite{Peter09}. With an incoherent (self luminous) screen there is very low spatial coherence and the method is essentially limited by photon noise. The ultimate limit of the height uncertainty however, depends on the lateral resolution as well, see Tab.~\ref{tab:uncertainty}. An interesting coupling of the angular uncertainty $\delta \alpha$ and the lateral resolution $\delta x$ leads to the useful uncertainty product given in Eq.~\ref{eq:defl}. The coupling is caused by the fact that the camera has to acquire  the object surface \textit{and} the remote screen pattern (sinusoidal fringes) simultaneously, which leads to a trade-off between angular uncertainty and lateral resolution \cite{Faber12}. More deeply, it is the Heisenberg uncertainty product that does not allow both, a small $\delta x$ and at the same time a small  $\delta \alpha $, for a single photon, but with many photons the SNR and the height resolution are virtually unlimited. Together with the signal-to-noise ratio SNR we get: 

\begin{equation}
    \delta z \approx \delta x \delta \alpha \approx 1/SNR
    \label{eq:defl}
\end{equation}

As a SNR=500 can easily be achieved with standard video cameras, a depth precision of $\delta z = 1 nm$ is possible with a few-dollar-equipment. Similar considerations prevail to classical shearing interferometry, but this requires more costly equipment and does not deliver a true derivative. The 'scaling' of shearing interferometry is given by the wavelength of light, while the scaling of  incoherent deflectometry is given by some macrosopic fringe generator. Nevertheless, incoherent deflectometry obviously can easily compete \cite{Faber12} and both methods are limited just by photon noise. 

Equation~\ref{eq:defl} offers numerous options: if we allow for a very low spatial resolution, for example to measure a big flat mirror, the angular uncertainty can be in the micro-arcsec range \cite{ehret12}.

For completeness we mention the well established Hartmann-Shack sensor as one more incoherent slope measuring sensor. It allows for a direct measurement of wave fronts coming for example from the pupil of a lens under test \cite{Hartmann04}. A certain limitation is the sampling of the wavefront by an array of discrete lenslets. 

Coherent differentiation is possible as well: Among the approximately “differentiating” methods there is shearing-interferometry and shearing-holography \cite{servin96, falldorf13} including its numerous (more or less coherent) implementations such as the Nomarski differential interference contrast \cite{Nomarski55}.
 
We move on to holographic shearing-interferometry, as it offers the possibility to measure the local slope of a \textit{rough} surface. A wave front is compared with a version of itself shifted (‘sheared’) by a distance $s$. A precious advantage of shearing methods (via a shearing plate) is the common path geometry, being insensitive against environmental perturbations. Shearing holography is realized via a shearing plate and capturing several sheared interference patterns corresponding to different shears. From the interference patterns, a hologram is calculated and can be evaluated by numerical back propagation \cite{Falldorf15}.”

To understand the physical limit of rough surface shearing holography, we refer again to the example of a projected laser spot: \cite{GH_88, GH_88_S}. A laser spot with diameter $d_{spot}$ is projected onto the object. In the hologram plane, an objective speckle pattern with speckle diameter $d_{speckle} \approx \lambda z / d_{spot}$ is generated and superimposed with a laterally shifted copy of itself. As far as the shear s is smaller than the speckle diameter, fringes can be seen, however with some phase noise, due to the phase decorrelation within each individual speckle. Earlier investigations reveal that this phase noise leads to a measuring uncertainty $\delta z \approx \lambda / \sin^2(u_{obs})$, which is just the Rayleigh depth of focus (why are we not surprised?). Obviously,  shearing interferometry at rough surfaces, as a tool to measure the local distance, displays (only) the same precision limit as focus searching methods. Hence it belongs to class I. This is a result which a naïve observer possibly would not have expected: Shearing holography at rough surfaces, to measure local distance, is equivalent to triangulation.

For an extended object and for a large shear, the situation is even more difficult. Principally, each surface point generates a fringe pattern in the hologram, but the phase is random, with respect to other surface points, which makes the deciphering of the hologram difficult. Nonetheless, the object can be fully reconstructed from the holograms \cite{Falldorf15}. The basic idea here is to determine the complex valued coherence function from the recorded interference patterns, starting from the mixed interference term $ E^*(x,y) E(x+s,y)$ of the complex signal $E$. From there, one can reconstruct finite differences of the wave field corresponding to positions separated by the shear, or by combining several shears, the non differentiated wave front. 

A real finite difference  can be achieved by using "$\Gamma$-profilometry", which exploits the temporal coherence function \cite{Falldorf21}. The method is strongly related to (scanning) rough-surface interferometry, as described in section~\ref{sec:RoughInt}. A significant difference to scanning white-light interferometry is that the outcome here is not the surface profile $z(x,y)$ but the difference $\Delta z = z(x,y)-z(x+s,y)$. This can be an advantage, as for most objects, the depth scanning time will be significantly lower, as the scanning range is limited by the maximum of $\Delta z$ instead by the full object depth range. This is an illustrative example for redundancy reduction via source encoding. It follows from these considerations that the ultimate source of noise for $\Gamma$-profilometry with broad-band illumination is given by the roughness of the surface: The method has to be assigned to class III, again with the useful feature that the measurement uncertainty does not depend on the observation aperture. It is obviously a fruitful idea, to "copy" and possibly to improve established methods like (scanning) rough surface interferometry by proper holographic storage and the big toolbox of computational evaluation.

Eventually, the so called 'shearography' \cite{francis10} has to be mentioned: Speckle shearography takes advantage from the common path geometry of shearing interferometry. The robustness against environmental perturbation is exploited to measure very small (temporal) changes of the surface “slope”, the sensitivity depending on the shear. The method is, of course, sensitive against speckle decorrelation. So only under the assumption of very small surface changes,  the precision might be limited by photon noise. For deformations close to a wavelength, phase decorrelation probably dominates.

\section{Holography vs. classical interferometry}

We have set aside this consideration until now, hoping that the foregoing sections simplify the understanding: As far as specular surfaces are involved, holography belongs to class II, classical interferometry, according to the fact that here the precision is limited by photon noise. 

One might however ask if the storage of a non-speckled wave front may be still called holography. As far as holography is defined just as "the storage and processing of a complex wave field", interferometry might go under the name of holography as well. The discussion is still open. Furthermore, we refer to the line of arguments in this paper.

Eventually, to holographic interferometry at rough objects: Here the object is compared with a slightly deformed version of itself. As discussed for speckle shearography, the phase of two highly correlated speckles is compared. As far as there is very low decorrelation, the precision is limited by photon noise, as for classical interferometry. This is however not a binary decision, as with increasing deformation, phase decorrelation and fringe delocalization occur.

\section{Conclusion}

Holography is a tool to store complex wave fronts and to process the data either optically or by computation. This allows for numerous implementations - amongst others - to acquire data about the topography of an object surface, which is the focus of this paper. The comparison of holographic methods with non-holographic methods offers some interesting insights. The first insight is that holographic methods can be assigned to one of four classes which are defined via the physical cause of the dominant noise and the dependence of the observation aperture. As not each and every method is investigated, the authors admit that further discussion might get new insight.

Methods based on triangulation are seriously disturbed by speckle, with the consequence that holography is not the first choice to measure surface topography via triangulation, e.g., by focus search or related methods. 3D-metrology, based on triangulation is the turf of incoherent methods, at least for \textit{macroscopic} objects. Many \textit{microscopical} objects are weak phase objects and there is no clear distinction of defocusing noise and speckle. Furthermore, the observation aperture can be extremely high - so by the help of Eq.~\ref{eq:dz_u}, holographic methods often deliver acceptable results, in spite of some coherent noise.

We think that holographic interferometry aiming to measure sub-$\lambda$ deformation, is the natural realm of holography, where most incoherent methods fail.

Two- or multi-wavelength holography, might have the potential  to become a competitor of the corresponding non-holographic methods that we find in the class "rough-surface interferometry". The potential of holography is strongly related to the vast realm of computational options. 'Seeing around the corner' is only one striking example, where several holographic ideas are combined.

Among the methods that intrinsically measure the slope, (incoherent) phase measuring deflectometry has developed to be an invaluable tool to measure virtually any kind of specular surfaces, even competing interferometry \cite{Faber12}.  Deflectometry displays an extreme sensitivity for local surface defects, with at the same time low hardware requirements. Can holography compete? Maybe for rough surfaces: Among the (approximately) differentiating holographic methods, multi-wavelength shearing-holography seems to be a proper method to measure the slope of rough surfaces. Again, the virtually unlimited options for computational post processing are an advantage over "purely" optical methods. The precision limit is determined just by the surface roughness, the same limit as for rough-surface interferometry. 

To conclude: Holography offers the presentation of breathtaking 3D images, but the underlying storage of a complex wave field, together with virtually unlimited options for processing by computation, is breathtaking as well. As there will be many not yet invented algorithms, to exploit holography,  future researchers might profit from knowing and exploiting the fundamental limits.

\begin{acknowledgements}
\noindent
\noindent
    The authors gratefully acknowledge numerous intriguing discussions with Claas Falldorf (BIAS Bremen) and Christian Faber (Hochschule Landshut), and as well their critical reading of the manuscript. 

\end{acknowledgements}

\section*{Bibliography}
\bibliography{zHenriquesLab-Mendeley}

\begin{thebibliography}{58}
\providecommand{\natexlab}[1]{#1}
\providecommand{\url}[1]{\texttt{#1}}
\expandafter\ifx\csname urlstyle\endcsname\relax
  \providecommand{\doi}[1]{doi: #1}\else
  \providecommand{\doi}{doi: \begingroup \urlstyle{rm}\Url}\fi

\bibitem[Leith and Upatnieks(1964)]{Leith_64}
Emmett~N. Leith and Juris Upatnieks.
\newblock Wavefront reconstruction with diffused illumination and
  three-dimensional objects$\ast$.
\newblock \emph{J. Opt. Soc. Am.}, 54\penalty0 (11):\penalty0 1295--1301, Nov
  1964.
\newblock \doi{10.1364/JOSA.54.001295}.

\bibitem[Powell and Stetson(1965)]{Powell_65}
Robert~L. Powell and Karl~A. Stetson.
\newblock Interferometric vibration analysis by wavefront reconstruction.
\newblock \emph{J. Opt. Soc. Am.}, 55\penalty0 (12):\penalty0 1593--1598, Dec
  1965.
\newblock \doi{10.1364/JOSA.55.001593}.

\bibitem[Offner(1963)]{Offner_63}
Abe Offner.
\newblock A null corrector for paraboloidal mirrors.
\newblock \emph{Appl. Opt.}, 2\penalty0 (2):\penalty0 153--155, Feb 1963.
\newblock \doi{10.1364/AO.2.000153}.

\bibitem[Lohmann and Paris(1967)]{Lohmann:67}
A.~W. Lohmann and D.~P. Paris.
\newblock Binary fraunhofer holograms, generated by computer.
\newblock \emph{Appl. Opt.}, 6\penalty0 (10):\penalty0 1739--1748, Oct 1967.
\newblock \doi{10.1364/AO.6.001739}.

\bibitem[Zhang and Yamaguchi(1998)]{Zhang:98}
Tong Zhang and Ichirou Yamaguchi.
\newblock Three-dimensional microscopy with phase-shifting digital holography.
\newblock \emph{Opt. Lett.}, 23\penalty0 (15):\penalty0 1221--1223, Aug 1998.
\newblock \doi{10.1364/OL.23.001221}.

\bibitem[Kim(2010)]{Kim_10}
Myung~K. Kim.
\newblock {Principles and techniques of digital holographic microscopy}.
\newblock \emph{SPIE Reviews}, 1\penalty0 (1):\penalty0 1 -- 51, 2010.
\newblock \doi{10.1117/6.0000006}.

\bibitem[Kemper et~al.(2007)Kemper, Langehanenberg, and von Bally]{Kemper_07}
Björn Kemper, Patrik Langehanenberg, and Gert von Bally.
\newblock Digital holographic microscopy.
\newblock \emph{Optik \& Photonik}, 2\penalty0 (2):\penalty0 41--44, 2007.
\newblock \doi{https://doi.org/10.1002/opph.201190249}.

\bibitem[H{\"a}usler and Ettl(2011)]{Haeusler_11}
Gerd H{\"a}usler and Svenja Ettl.
\newblock \emph{Limitations of Optical 3D Sensors}.
\newblock Richard Leach (Ed), Optical Measurement of Surface Topography, pages
  23--48, Springer Berlin Heidelberg, 2011.
\newblock ISBN 978-3-642-12012-1.
\newblock \doi{10.1007/978-3-642-12012-1_3}.

\bibitem[H\"{a}usler(2019)]{Haeusler_19}
Gerd H\"{a}usler.
\newblock Discover better optical sensors - by exploring and exploiting
  nature's limits.
\newblock In \emph{Imaging and Applied Optics 2019 (COSI, IS, MATH, pcAOP)},
  page CTh2A.1. Optical Society of America, 2019.
\newblock \doi{10.1364/COSI.2019.CTh2A.1}.

\bibitem[Srinivasan et~al.(1984)Srinivasan, Liu, and Halioua]{Srinivasan_84}
V.~Srinivasan, H.~C. Liu, and M.~Halioua.
\newblock Automated phase-measuring profilometry of 3-d diffuse objects.
\newblock \emph{Appl. Opt.}, 23\penalty0 (18):\penalty0 3105--3108, Sep 1984.
\newblock \doi{10.1364/AO.23.003105}.

\bibitem[Takeda and Mutoh(1983)]{Takeda_83}
Mitsuo Takeda and Kazuhiro Mutoh.
\newblock Fourier transform profilometry for the automatic measurement of 3-d
  object shapes.
\newblock \emph{Appl. Opt.}, 22\penalty0 (24):\penalty0 3977--3982, Dec 1983.
\newblock \doi{10.1364/AO.22.003977}.

\bibitem[Willomitzer and H\"{a}usler(2017)]{Willo_17}
F.~Willomitzer and G.~H\"{a}usler.
\newblock Single-shot 3d motion picture camera with a dense point cloud.
\newblock \emph{Opt. Express}, 25\penalty0 (19):\penalty0 23451--23464, Sep
  2017.
\newblock \doi{10.1364/OE.25.023451}.

\bibitem[Huber et~al.(2011)Huber, Arold, Willomitzer, Ettl, and
  H{\"a}usler]{Huber_13}
F.~Huber, O.~Arold, F.~Willomitzer, S.~Ettl, and G.~H{\"a}usler.
\newblock 3d body scanning with flying triangulation.
\newblock In \emph{Proceedings of the 112th DGaO Conference, P30}, 2011.

\bibitem[Takeda et~al.(1982)Takeda, Ina, and Kobayashi]{Takeda_82}
Mitsuo Takeda, Hideki Ina, and Seiji Kobayashi.
\newblock Fourier-transform method of fringe-pattern analysis for
  computer-based topography and interferometry.
\newblock \emph{J. Opt. Soc. Am.}, 72\penalty0 (1):\penalty0 156--160, Jan
  1982.
\newblock \doi{10.1364/JOSA.72.000156}.

\bibitem[H{\"a}usler et~al.(2012)H{\"a}usler, Faber, Willomitzer, and
  Dienstbier]{GH_12}
G.~H{\"a}usler, C.~Faber, F.~Willomitzer, and P.~Dienstbier.
\newblock Why can’t we purchase a perfect single shot 3d-sensor?
\newblock In \emph{Proceedings of the 113th DGaO Conference, A8}, 2012.

\bibitem[Bruning et~al.(1974)Bruning, Herriott, Gallagher, Rosenfeld, White,
  and Brangaccio]{Bruning_74}
J.~H. Bruning, D.~R. Herriott, J.~E. Gallagher, D.~P. Rosenfeld, A.~D. White,
  and D.~J. Brangaccio.
\newblock Digital wavefront measuring interferometer for testing optical
  surfaces and lenses.
\newblock \emph{Appl. Opt.}, 13\penalty0 (11):\penalty0 2693--2703, Nov 1974.
\newblock \doi{10.1364/AO.13.002693}.

\bibitem[Hol(Retrieved 12/05/2021)]{HoloImage}
Image taken from
  \url{https://commons.wikimedia.org/wiki/File:Holographic_recording.jpg},
  Retrieved 12/05/2021.

\bibitem[Willomitzer(2019)]{Willo_19Diss}
F.~Willomitzer.
\newblock \emph{Single-Shot 3D Sensing Close to Physical Limits and Information
  Limits}.
\newblock Springer Theses, 2019.

\bibitem[Willomitzer et~al.(2015)Willomitzer, Ettl, Faber, and
  H{\"a}usler]{Willo_15}
F.~Willomitzer, S.~Ettl, C.~Faber, and G.~H{\"a}usler.
\newblock Single-shot three-dimensional sensing with improved data density.
\newblock \emph{Applied Optics}, 54\penalty0 (3):\penalty0 408--417, 2015.

\bibitem[Cheng and Wyant(1984)]{Cheng_84}
Yeou-Yen Cheng and James~C. Wyant.
\newblock Two-wavelength phase shifting interferometry.
\newblock \emph{Appl. Opt.}, 23\penalty0 (24):\penalty0 4539--4543, Dec 1984.
\newblock \doi{10.1364/AO.23.004539}.

\bibitem[Falaggis et~al.(2011)Falaggis, Towers, and Towers]{Falaggis_11}
Konstantinos Falaggis, David~P. Towers, and Catherine~E. Towers.
\newblock Method of excess fractions with application to absolute distance
  metrology: theoretical analysis.
\newblock \emph{Appl. Opt.}, 50\penalty0 (28):\penalty0 5484--5498, Oct 2011.
\newblock \doi{10.1364/AO.50.005484}.

\bibitem[Dorsch et~al.(1994)Dorsch, H\"{a}usler, and Herrmann]{Dorsch_94}
Rainer~G. Dorsch, Gerd H\"{a}usler, and J\"{u}rgen~M. Herrmann.
\newblock Laser triangulation: fundamental uncertainty in distance measurement.
\newblock \emph{Appl. Opt.}, 33\penalty0 (7):\penalty0 1306--1314, Mar 1994.
\newblock \doi{10.1364/AO.33.001306}.

\bibitem[H{\"a}usler(Encyclopedia of Modern Optics, Elsevier, Academic Press.,
  Oxford, pp 114-123, 2004)]{GH_04_Enc}
G.~H{\"a}usler.
\newblock Speckle and coherence, Encyclopedia of Modern Optics, Elsevier,
  Academic Press., Oxford, pp 114-123, 2004.

\bibitem[Wagner and H{\"a}usler(2003)]{Wagner_03}
Christoph Wagner and Gerd H{\"a}usler.
\newblock Information theoretical optimization for optical range sensors.
\newblock \emph{Applied optics}, 42\penalty0 (27):\penalty0 5418--5426, 2003.

\bibitem[H{\"a}usler et~al.(1988)H{\"a}usler, Hutfless, Maul, and
  Weissmann]{GH_88}
Gerd H{\"a}usler, Jochen Hutfless, Manfred Maul, and Hans Weissmann.
\newblock Range sensing based on shearing interferometry.
\newblock \emph{Applied Optics}, 27\penalty0 (22):\penalty0 4638--4644, 1988.

\bibitem[H{\"a}usler and Herrmann(1988)]{GH_88_S}
G.~H{\"a}usler and J.{\"u}rgen~M. Herrmann.
\newblock Range sensing by shearing interferometry: influence of speckle.
\newblock \emph{Applied optics}, 27\penalty0 (22):\penalty0 4631--4637, 1988.

\bibitem[Thorley et~al.(2014)Thorley, Pike, and Rappoport]{Thorley_14}
Jennifer~A Thorley, Jeremy Pike, and Joshua~Z Rappoport.
\newblock Super-resolution microscopy: a comparison of commercially available
  options.
\newblock In \emph{Fluorescence Microscopy}, pages 199--212. Elsevier, 2014.

\bibitem[Spellenberg et~al.(2001)Spellenberg, Herrmann, and
  H{\"a}usler]{spellenberg_01}
B.~Spellenberg, J.~M. Herrmann, and G.~H{\"a}usler.
\newblock Highly improved range sensing--by reduction of spatial coherence.
\newblock \emph{Optik}, 112\penalty0 (7):\penalty0 299--303, 2001.

\bibitem[Dresel et~al.(1992)Dresel, H{\"a}usler, and Venzke]{Dresel92}
Thomas Dresel, Gerd H{\"a}usler, and Holger Venzke.
\newblock Three-dimensional sensing of rough surfaces by coherence radar.
\newblock \emph{Applied optics}, 31\penalty0 (7):\penalty0 919--925, 1992.

\bibitem[Caber(1993)]{Caber93}
Paul~J Caber.
\newblock Interferometric profiler for rough surfaces.
\newblock \emph{Applied optics}, 32\penalty0 (19):\penalty0 3438--3441, 1993.

\bibitem[Fercher et~al.(1985)Fercher, Hu, and Vry]{fercher85}
A.F. Fercher, Hong~Zhang Hu, and U.~Vry.
\newblock Rough surface interferometry with a two-wavelength heterodyne speckle
  interferometer.
\newblock \emph{Applied optics}, 24\penalty0 (14):\penalty0 2181--2188, 1985.

\bibitem[Vry and Fercher(1986)]{vry86}
U~Vry and AF~Fercher.
\newblock Higher-order statistical properties of speckle fields and their
  application to rough-surface interferometry.
\newblock \emph{JOSA A}, 3\penalty0 (7):\penalty0 988--1000, 1986.

\bibitem[D{\"a}ndliker et~al.(1988)D{\"a}ndliker, Thalmann, and
  Prongu{\'e}]{dandliker88}
Ren{\'e} D{\"a}ndliker, R~Thalmann, and D~Prongu{\'e}.
\newblock Two-wavelength laser interferometry using superheterodyne detection.
\newblock \emph{Optics letters}, 13\penalty0 (5):\penalty0 339--341, 1988.

\bibitem[Li et~al.(2018)Li, Willomitzer, Rangarajan, Gupta, Velten, and
  Cossairt]{Li18}
F.~Li, F.~Willomitzer, P.~Rangarajan, M.~Gupta, A.~Velten, and O.~Cossairt.
\newblock Sh-tof: Micro resolution time-of-flight imaging with superheterodyne
  interferometry.
\newblock In \emph{2018 IEEE International Conference on Computational
  Photography (ICCP)}, pages 1--10. IEEE, 2018.

\bibitem[Wagner et~al.(2000)Wagner, Osten, and Seebacher]{Wagner00}
Chr. Wagner, Wolfgang Osten, and Soenke Seebacher.
\newblock {Direct shape measurement by digital wavefront reconstruction and
  multi-wavelength contouring}.
\newblock \emph{Optical Engineering}, 39\penalty0 (1):\penalty0 79 -- 85, 2000.
\newblock \doi{10.1117/1.602338}.

\bibitem[Willomitzer et~al.(2021)Willomitzer, Rangarajan, Li, Balaji,
  Christensen, and Cossairt]{Willo21}
F.~Willomitzer, P.~V. Rangarajan, F.~Li, M.~M. Balaji, M.~P. Christensen, and
  O.~Cossairt.
\newblock Fast non-line-of-sight imaging with high-resolution and wide field of
  view using synthetic wavelength holography.
\newblock \emph{Nature communications}, 12\penalty0 (1):\penalty0 1--11, 2021.

\bibitem[Willomitzer et~al.(2019{\natexlab{a}})Willomitzer, Li, Balaji,
  Rangarajan, and Cossairt]{Willo19COSI}
F.~Willomitzer, F.~Li, M.~M. Balaji, P.~Rangarajan, and O.~Cossairt.
\newblock High resolution non-line-of-sight imaging with superheterodyne remote
  digital holography.
\newblock In \emph{Imaging and Applied Optics, Computational Optical Sensing
  and Imaging}, pages CM2A--2. Optical Society of America, 2019{\natexlab{a}}.

\bibitem[Willomitzer et~al.(2019{\natexlab{b}})Willomitzer, Rangarajan, Li,
  Balaji, Christensen, and Cossairt]{Willo19Arx}
F.~Willomitzer, P.~V. Rangarajan, F.~Li, M.~M. Balaji, M.~P. Christensen, and
  O.~Cossairt.
\newblock Synthetic wavelength holography: An extension of gabor's holographic
  principle to imaging with scattered wavefronts.
\newblock \emph{arXiv preprint arXiv:1912.11438}, 2019{\natexlab{b}}.

\bibitem[Li et~al.(2021)Li, Willomitzer, Balaji, Rangarajan, and
  Cossairt]{Li21}
F.~Li, F.~Willomitzer, M.~M. Balaji, P.~Rangarajan, and O.~Cossairt.
\newblock Exploiting wavelength diversity for high resolution time-of-flight 3d
  imaging.
\newblock \emph{IEEE Transactions on Pattern Analysis and Machine
  Intelligence}, 2021.

\bibitem[George et~al.(1975)George, Jain, and Melville]{George75}
Nicholas George, Atul Jain, and RDS Melville.
\newblock Experiments on the space and wavelength dependence of speckle.
\newblock \emph{Applied physics}, 7\penalty0 (3):\penalty0 157--169, 1975.

\bibitem[Ettl(Diploma Thesis, University Erlangen-Nuremberg 1995)]{Ettl95}
Peter Ettl.
\newblock \emph{Studien zur hochgenauen Objektvermessung mit dem
  Koh{\"a}renzradar"}.
\newblock Diploma Thesis, University Erlangen-Nuremberg 1995.

\bibitem[H{\"a}usler et~al.(1999)H{\"a}usler, Ettl, Schenk, Bohn, and
  Laszlo]{GH99Asa}
Gerd H{\"a}usler, Peter Ettl, M~Schenk, Gunther Bohn, and Ildiko Laszlo.
\newblock Limits of optical range sensors and how to exploit them.
\newblock In \emph{International trends in optics and photonics}, pages
  328--342. Springer, 1999.

\bibitem[Kwon et~al.(1980)Kwon, Wyant, and Hayslett]{kwon80}
Osuk Kwon, JC~Wyant, and CR~Hayslett.
\newblock Rough surface interferometry at 10.6 $\mu$m.
\newblock \emph{Applied optics}, 19\penalty0 (11):\penalty0 1862--1869, 1980.

\bibitem[H{\"a}usler(1999)]{GHDefPat}
G.~H{\"a}usler.
\newblock \emph{Verfahren und Vorrichtung zur Ermittlung der Form oder der
  Abbildungseigenschaften von spiegelnden oder transparenten Objekten, Patent
  DE}, 19944354:\penalty0 A1, 1999.

\bibitem[Knauer et~al.(2004)Knauer, Kaminski, and H{\"a}usler]{Knauer04}
Markus~C Knauer, Jurgen Kaminski, and Gerd H{\"a}usler.
\newblock Phase measuring deflectometry: a new approach to measure specular
  free-form surfaces.
\newblock In \emph{Optical Metrology in Production Engineering}, volume 5457,
  pages 366--376. International Society for Optics and Photonics, 2004.

\bibitem[Willomitzer et~al.(2020)Willomitzer, Yeh, Gupta, Spies, Schiffers,
  Katsaggelos, Walton, and Cossairt]{Willo20Def}
F.~Willomitzer, C-K Yeh, V.~Gupta, W.~Spies, F.~Schiffers, A.~Katsaggelos,
  M.~Walton, and O.~Cossairt.
\newblock Hand-guided qualitative deflectometry with a mobile device.
\newblock \emph{Optics express}, 28\penalty0 (7):\penalty0 9027--9038, 2020.

\bibitem[Bates(Nature, 158, 221 (1946).)]{Bates46}
W.J. Bates, Nature, 158, 221 (1946).

\bibitem[H{\"a}usler et~al.(2008)H{\"a}usler, Richter, Leitz, and
  Knauer]{GH08MD}
Gerd H{\"a}usler, Claus Richter, Karl-Heinz Leitz, and Markus~C Knauer.
\newblock Microdeflectometry—a novel tool to acquire three-dimensional
  microtopography with nanometer height resolution.
\newblock \emph{Optics letters}, 33\penalty0 (4):\penalty0 396--398, 2008.

\bibitem[Peterhänsel et~al.(2009)Peterhänsel, Richter, Faber, Knauer, and
  H{\"a}usler]{Peter09}
S.~Peterhänsel, C.~Richter, C.~Faber, M.~C. Knauer, and G.~H{\"a}usler.
\newblock Microdeflectometry in transmission.
\newblock In \emph{in Proceedings of the 110th DGaO Conference, P24}, 2009.

\bibitem[Faber et~al.(2012)Faber, Olesch, Krobot, and H{\"a}usler]{Faber12}
Christian Faber, Evelyn Olesch, Roman Krobot, and Gerd H{\"a}usler.
\newblock Deflectometry challenges interferometry: the competition gets
  tougher!
\newblock In \emph{Interferometry XVI: Techniques and Analysis}, volume 8493,
  page 84930R. International Society for Optics and Photonics, 2012.

\bibitem[Ehret et~al.(2012)Ehret, Schulz, Stavridis, and Elster]{ehret12}
G.~Ehret, M.~Schulz, M.~Stavridis, and C.~Elster.
\newblock Deflectometric systems for absolute flatness measurements at ptb.
\newblock \emph{Measurement Science and Technology}, 23\penalty0 (9):\penalty0
  094007, 2012.

\bibitem[Hartmann(Z.f. Instrumentenkunde 24, 1–21, 1904)]{Hartmann04}
J.~Hartmann.
\newblock Objektivuntersuchungen, Z.f. Instrumentenkunde 24, 1–21, 1904.

\bibitem[Servin et~al.(1996)Servin, Malacara, and Marroquin]{servin96}
M~Servin, Daniel Malacara, and JL~Marroquin.
\newblock Wave-front recovery from two orthogonal sheared interferograms.
\newblock \emph{Applied optics}, 35\penalty0 (22):\penalty0 4343--4348, 1996.

\bibitem[Falldorf et~al.(2013)Falldorf, von Kopylow, and Bergmann]{falldorf13}
Claas Falldorf, Christoph von Kopylow, and Ralf~B Bergmann.
\newblock Wave field sensing by means of computational shear interferometry.
\newblock \emph{JOSA A}, 30\penalty0 (10):\penalty0 1905--1912, 2013.

\bibitem[Nomarski(1955)]{Nomarski55}
G.~M. Nomarski.
\newblock Differential microinterferometer with polarized waves.
\newblock \emph{J. Phys. Radium Paris}, 16:\penalty0 9S, 1955.

\bibitem[Falldorf et~al.(2015)Falldorf, Agour, and Bergmann]{Falldorf15}
Claas Falldorf, Mostafa Agour, and Ralf~B Bergmann.
\newblock Digital holography and quantitative phase contrast imaging using
  computational shear interferometry.
\newblock \emph{Optical Engineering}, 54\penalty0 (2):\penalty0 024110, 2015.

\bibitem[Falldorf et~al.(2021)Falldorf, Agour, M{\"u}ller, and
  Bergmann]{Falldorf21}
Claas Falldorf, Mostafa Agour, Andr{\'e}~F M{\"u}ller, and Ralf~B Bergmann.
\newblock $\gamma$-profilometry: a new paradigm for precise optical metrology.
\newblock \emph{Optics Express}, 29\penalty0 (22):\penalty0 36100--36110, 2021.

\bibitem[Francis et~al.(2010)Francis, Tatam, and Groves]{francis10}
Daniel Francis, RP~Tatam, and RM~Groves.
\newblock Shearography technology and applications: a review.
\newblock \emph{Measurement science and technology}, 21\penalty0 (10):\penalty0
  102001, 2010.

\end{thebibliography}


\end{document}